\documentclass{svjour3}                    % onecolumn
\usepackage{graphicx}
\usepackage{epstopdf}
\begin{document}

\title{Characterization of the Surface of Moving Solid $^4$He
%\thanks{This work was supported by The Israel Science Foundation, grant 1089/13 and by the Technion Fund for Research }
}


\author{Ethan Livne         \and
        Anna Eyal			\and
        Ori Scaly			\and
		Emil Polturak		%etc.
}


\institute{E. Livne \at
              Department of Physics, Technion - Israel Institute of Technology, Haifa 32000, Israel\\
			  %\email{livne@physics.technion.ac.il}
			\and
           A. Eyal \at
              Laboratory of Atomic and Solid State Physics, Cornell University, Ithaca, NY 14853, USA
			\and
           O. Scaly \at
              Department of Physics, Technion - Israel Institute of Technology, Haifa 32000, Israel\\
              \and
			E. Polturak \at
			Department of Physics, Technion - Israel Institute of Technology, Haifa 32000, Israel
              Tel.: +972-8-8292761\\
              Fax: +972-8-8292027\\
			\\\email{emilp@physics.technion.ac.il}
}

\date{Received: date / Accepted: date}

\maketitle

\begin{abstract}
Crystal grains of solid $^4$He can move in relation to each other even
when embedded inside the solid \cite{a001,a002}. In this work, we
characterize a macroscopic motion of solid hcp $^4$He composed of such
grains. Motion is induced by applying an external torque to the solid
contained inside an annular channel mounted on a torsional oscillator.
In order to characterize the surface of the moving solid, we developed
an in-situ flow detection method using a sensitive ``microphone''
embedded in the wall of the channel. Motion is detected by counting the
vibrations induced by rows of He atoms moving past the microphone. Such
vibrations were detected only at T=0.5K, our lowest temperature. At
this temperature, the measured dissipation associated with the solid He
is zero within our accuracy. Our results indicate that the orientation
of the surface of the moving solid is the (0001) basal plane of the hcp
structure.  At T=0.5 K, we found that for speeds $<7\,\mu$m/sec, the solid
flows without detectable friction.

\keywords{Solid Helium \and Friction}
\end{abstract}

\section{Introduction}
\label{intro}

Solid $^4$He, the archetypal quantum solid, has been the subject of
intense study in recent years. Aside from a search for supersolidity
\cite{a003}, the existence of which is currently under debate \cite{a004,a005,a006},
there is an ongoing effort to explore and
understand the unique elastic \cite{a007} and
plastic \cite{a008,a009,a010} properties of crystalline He
at very low temperatures.
We are particularly interested in understanding the friction between
crystallites of He moving past each other. Internal friction in a
static polycrystalline solid He was investigated in the past, and was
found to result from dynamics of pinned dislocations as described by
the Lucke-Granato model \cite{a011,a012}. However, the classical
friction problem of one solid mass moving against another, was not yet
addressed in solid He. This is an interesting issue, since beyond
classical friction, there are several predictions regarding what the
friction would be in the quantum regime. One proposed friction
mechanism comes from the Van der Waals force, due to zero point charge
fluctuations \cite{a013}. Another mechanism involves phonon-phonon
interaction while the phonons populating one of the masses are Doppler
shifted due to the relative motion \cite{a014}. Since solid He exists
only under pressure, the classical layout of a tabletop friction
experiment cannot be realized. We have however accidentally stumbled
upon a way to perform such friction experiments in the way described
below. Preliminary results \cite{a015} indicate that the friction
between two He crystallites is extremely low, yet it can be measured
with good precision. We hope that such experiments may pave the way to
future search for fundamental mechanisms of quantum friction. The
experiment described here is a first step in this direction, intended
to characterize the interface between two moving masses of solid He.
The paper is organized as follows: We first describe our crystal
preparation. This subject is similar to work discussed previously and
is included here to make the paper self contained. The second part
describes our new in-situ motion detection method. The third and final
part includes the experimental results and discussion thereof.

\section{Experimental I: Growth and disordering of $^4$He crystals}
\label{experimental1}

Our interest in doing the work discussed here began with neutron
diffraction experiments on bcc solid $^4$He. During these experiments % fixed typo (EL)
we grew large (7-10\,cm$^3$) single crystals, which sometimes
transformed into polycrystals due to some thermal or mechanical stress.
In those cases, we noticed that individual crystallites which are part
of a polycrystalline solid can spontaneously change their spatial
orientation while embedded in the surrounding solid \cite{a001}. These
orientation changes seemed to be driven by ambient noise, either
vibrational or thermal. Consequently, these orientation changes of the
solid grains persisted for the duration of the experiment (on an order
of a week). Usual solids deform under applied shear stress. Seemingly
spontaneous motion of macroscopic solid crystallites inside a solid
matrix is unusual, and seems to be unique to solid He. The same unusual
behavior was later observed \cite{a002} in polycrystalline hcp $^4$He.
In contrast to polycrystals, single crystals of solid He are stable in
the sense that they do not change their orientation at all. In the
neutron scattering setup, the cell containing solid He was static and
the ambient stress field was caused by mechanical vibrations or thermal
noise. To see how this spontaneous mobility of the polycrystalline
solid manifests itself under a controlled external stress, we decided
to study similar crystals grown inside an annular channel forming a
part of a torsional oscillator (TO). In this setting, the solid is
acted upon by an external torque applied by the moving walls of the TO.

\begin{figure} % Figure 1 - TO
  \includegraphics[width=1\textwidth]{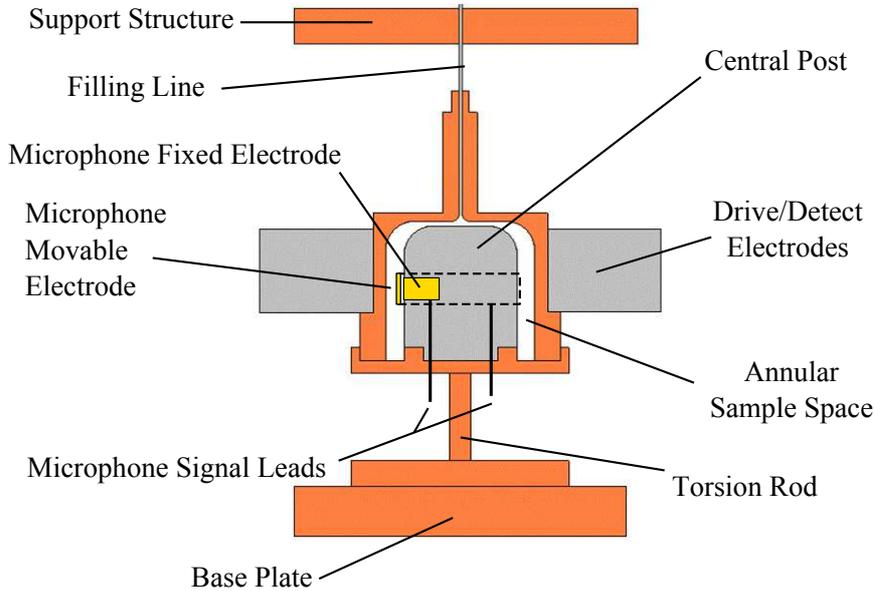}

	\caption{
		A schematic cross sectional drawing of the TO used in this work. The annular sample cell, of 2\,mm width and 10\,mm height is filled with He through the heated filling line of 0.25\,mm i.d. The motion sensor(microphone) is part of the central post. Its structure and properties are discussed in the text.
	}
	\label{fig:TO}
\end{figure}

\begin{figure} % Figure 2 - break
  \includegraphics[width=1\textwidth]{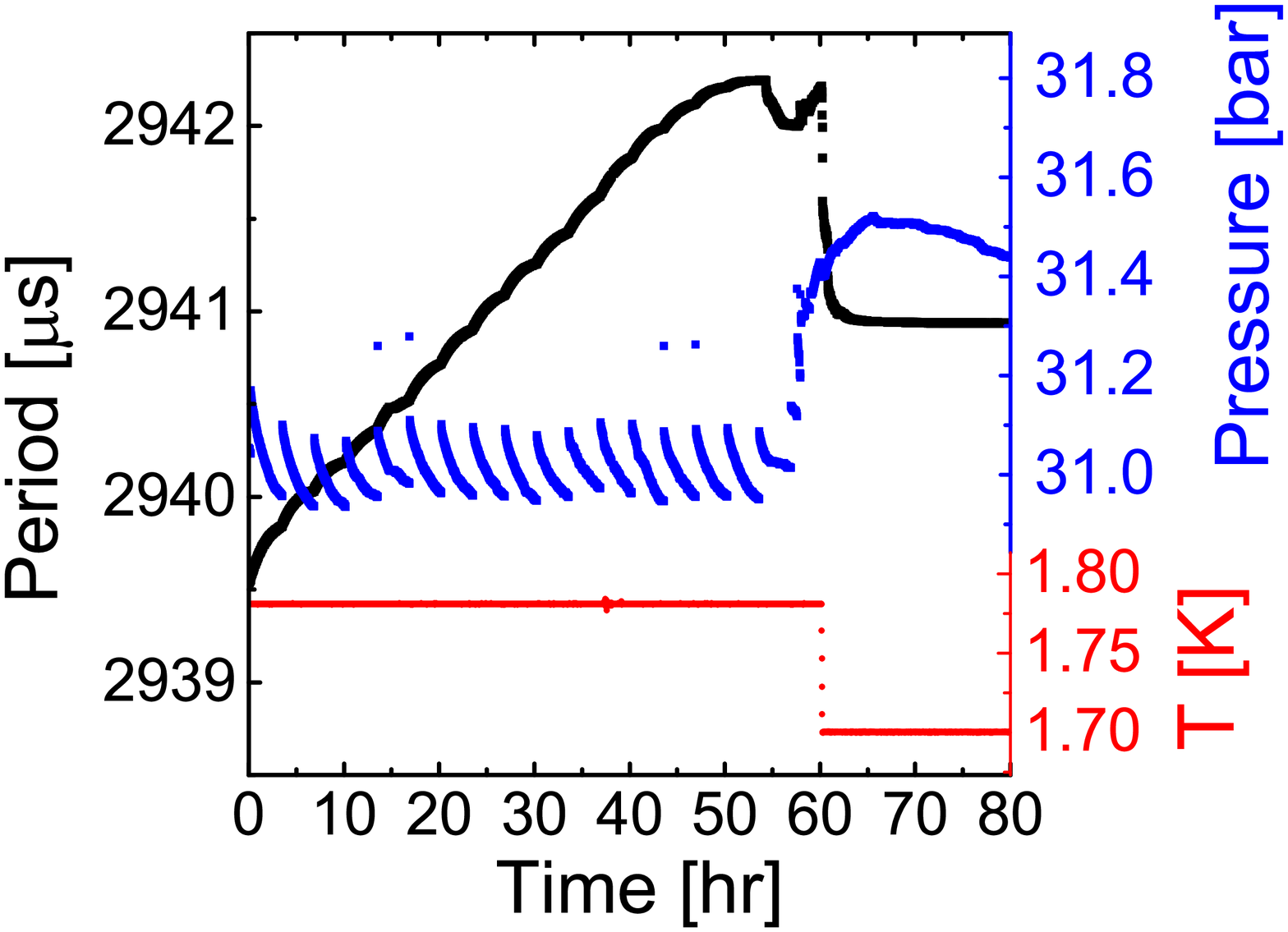}

	\caption{
		Monitor data during final stages of growth and subsequent disordering of a single hcp crystal grown on the melting curve at a temperature close to 1.8K. Time is measured from the beginning of crystal growth. During growth, small quantities of He are added periodically to the cell through a heated fill line, causing the pressure (blue symbols, right axis) to increase momentarily and then relax back to the melting pressure as liquid is converted into solid. Simultaneously, the resonant period of the TO (black symbols) increases as additional solid mass is coupled to the oscillator. After about 55 hours, the cell is full of solid and the pressure is increased further to block the filling line with solid. The heat coming through the filling line is cut off and the crystal begins to disorder. At 60.2 hours, the temperature (red symbols) of the cell is lowered to 1.7K. This causes the crystal to disorder further and the period of the TO spontaneously to decrease, as if some of the solid becomes decoupled from the motion of the cell. This is what we refer to as ``mass decoupling''.
	}
	\label{fig:break}
\end{figure}

We grow single crystals of solid hcp $^4$He of commercial purity %Removed a redundant space (EL)
(0.3\,ppm $^3$He) inside an annular channel forming part of the
TO \cite{a017,a018}, shown in Figure \ref{fig:TO}. The channel width is
2\,mm, and its height is 10\,mm. The channel is filled through a heated
filling line, 0.25\,mm i.d., connected to the top of the cell. Heating
the filling line ensures that the lowest temperature is at the bottom
of the cell, nearest to the cold plate, and so the solid nucleates
there. We use the same growth method developed during the neutron
scattering experiments. Crystal growth takes place at a constant
temperature and pressure, on the melting curve, by periodically adding
small amounts of helium into the cell. The TO is thermally attached to
a $^3$He cryostat which provides a stable platform at any temperature
between 0.5\,K and 3\,K.  The measured contribution of a single crystal of
$^4$He to the moment of inertia of the TO agrees very well with the
calculated value. There is no increase in the internal dissipation of
the TO associated with a single crystal. Hence, within our accuracy a
single crystal behaves like a perfect rigid body moving with the TO.
Figure \ref{fig:break} shows how the experiment is monitored during
final stages of crystal growth. Once the cell is full of solid, we
block the filling line with solid as well. Next, we stress the crystal
further by cooling the cell by 50-100\,mK. Thermal contraction of solid
He is prevented by the central post of the annular sample space. As a
result, internal stress is generated, large enough to cause the crystal
to disorder and transform into a polycrystal.  This behavior is
consistent with what we observed in neutron scattering
experiments \cite{a001}.  It is important to point out that the crystal can be disordered
only after the cell is entirely filled with solid. As long as any
liquid remains in the cell, it acts as a buffer which relieves stress,
and the single crystal is stable. Even with one drop of liquid in the
cell, we were able to maintain single crystals for weeks on end. On the
other hand, once the cell is full of solid, there is no such buffer. In
this situation, temperature changes induce stress which causes the
single crystal to disintegrate.

\begin{figure} % Figure 3 qfactor
  \begin{center}
  \includegraphics[width=.6\textwidth]{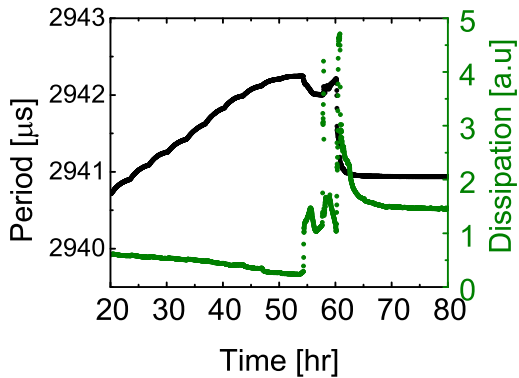}
  \end{center}

	\caption{
		Time dependence of the period and dissipation of the TO during final stages of crystal growth and subsequent disordering.
	}
	\label{fig:qfactor}
\end{figure}

Figure \ref{fig:qfactor} shows in more detail how the transformation of
a single crystal to a polycrystal affects the TO. Once the disordering
is complete (time $>$ 60 hours), the resonant period (black symbols) has decreased and
in parallel, the dissipation of the TO (black symbols) increased.
First, we discuss the resonant period. A decrease of the period of the
TO can be caused by several reasons: (a) partial melting of the solid.
We start to cool with both the cell and its filling line already full
of solid. Cooling, if anything, converts liquid into solid, not the
other way round. Hence, option (a) is not possible. We checked this
directly by intentionally introducing some liquid into the cell. The
polycrystal immediately annealed, the period of the TO returned to that
of a cell filled with a single crystal, and the decoupling effect
vanished.  Option (b) is that the resonant frequency increases due to
an increase of the effective torsion constant of the TO. This can
happen if the shear modulus of solid He increases. This
possibility was discussed by Maris and Balibar \cite{a016} in conjunction with the search for supersolidity\cite{a003}
at low temperatures. To see whether this possibility applies in our case,
we carried out a similar analysis of our system and for our temperature range using Finite Element Analysis.
This analysis is described in detail in the Appendix. The conclusion is that various effects related to the shear modulus of
solid He affect the resonant period of the TO by 1 to 2 orders of magnitude less than the changes we observe.
Consequently, effects related to the shear modulus cannot explain our results either. We are
therefore left with option (c), that the effective moment of inertia
associated with a polycrystal is smaller than that of a single crystal.
This will take place if part of the solid does not move as a rigid body
with the TO and becomes effectively decoupled from the motion of the
walls. In the frame of reference of the TO, this decoupled solid moves
relatively to the wall, in consistency with our observations during the
neutron scattering experiments. %removed redundant line break. (EL)

We now consider the dissipation of the TO, shown in Figure
\ref{fig:qfactor} as green symbols. The dissipation is given by
$\omega_{TO}\times\left(\mathrm{stored\, energy}\right)/2\pi Q$, where
$\omega_{TO}/2\pi$ is the resonant frequency of the TO and $Q$ is the
quality factor. At times $<$ 55 hours, while the cell is being filled
with a single crystal, the dissipation decreases. This indicates that
the single crystal behaves as a rigid body and the remnant dissipation
is associated with the fluid remaining in the cell. When the cell is
entirely filled with solid, the dissipation reaches a minimum. Upon
disordering the crystal, the dissipation increases, and after several
hours reaches a steady state. The increase of the dissipation indicates
that with a polycrystal, there is additional friction associated with
internal motion within the solid. Based on neutron diffraction
experiments \cite{a001}, we associate the time dependent changes of the
dissipation with orientation changes of the polycrystal inside the
cell.  The reorientation process gradually lowers the internal
dissipation from its peak value. To calculate the dissipation, we use
the moment of inertia of our TO which is about 40\,gm$\cdot$cm$^2$.
Taking a typical tangential velocity of 10\,$\mu$m/sec, the energy stored
in the TO is in the 10$^{-5}$\,erg range. We found that the dissipation
associated with the oriented polycrystal is temperature
dependent \cite{a015}. At temperatures above 1\,K, the dissipation is in
the 10$^{-7}$\,erg/sec range. At our lowest temperature of 0.5\,K, it
decreases to 10$^{-9}$\,erg/sec, which within our precision is the same
as that of an empty cell. Hence, at 0.5\,K the internal friction between
the decoupled solid and the solid moving with the wall is zero within
experimental accuracy. Systematic measurements of this dissipation can
tell us about the internal friction, a subject for our future work.

Finally, we point out the necessity of growing single crystals and
turning them into oriented polycrystals. For comparison, we grew
unoriented polycrystals inside the cell instead of single
crystals \cite{a019}. In this case, we observed ≤ 1\%  of mass
decoupling instead of 10\%-25\%. Hence, starting with a single crystal
is essential to maximize the mass decoupling. We also point out that
the transformation of a single crystal to a polycrystal occurs very
close to the temperature at which a particular crystal was grown. We
can grow single crystals and disorder them at any temperature of choice
between 1.1\,K and 2.5\,K, so this transformation is a thermomechanical
effect, not a phase transition. In particular, we see no apparent
connection between our work and the search for
supersolidity \cite{a004,a005,a006,a007} which is done
at much lower temperatures.

\section{Experimental II: Microphone study of the motion of solid $^4$He}
\label{experimental2}

\begin{figure} % Figure 4 working-model
  \begin{center}
  \includegraphics[width=.6\textwidth]{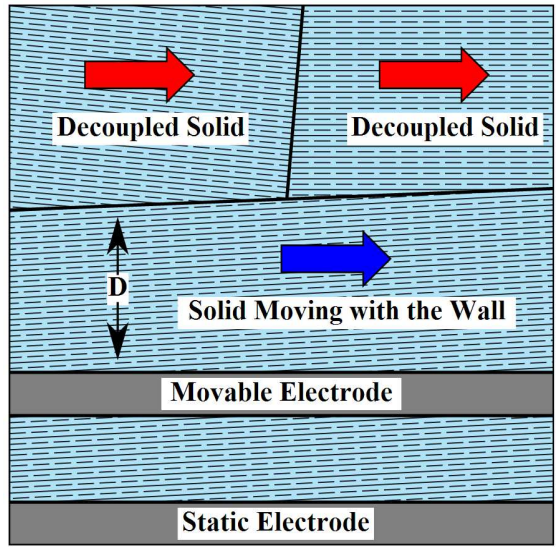}
  \end{center}

	\caption{
		Schematic cross-sectional illustration of our working model: the static electrode of the microphone is embedded in the wall of the TO(see Figure \ref{fig:TO}). A grain boundary at a distance $D$ from the movable electrode separates the solid moving with the wall and the decoupled solid. Blue and red arrows illustrate that the motion of the decoupled solid (red arrow) can differ from that of the wall and the static solid He near it (blue arrow). The amplitude of the motion is small, typically 5-15 lattice constants.
	}
	\label{fig:working-model}
\end{figure}

So far, the information regarding motion of the solid relative to the
walls of the TO was indirect, inferred from the change of the moment of
inertia of the TO. To learn more about what actually goes on inside the
cell, we developed a new method to study this motion in-situ.
Specifically, we designed a sensitive ``microphone'' which we
incorporated into the inner wall of the annular sample space inside the
TO (see Figure \ref{fig:TO}). Essentially, the microphone is a plate
capacitor with one plate which is movable. The static electrode is a
2mm diameter brass disc, embedded in the central post as shown in Fig
\ref{fig:TO}. The movable electrode is a 15\,$\mu$m thick metalized
polyvinyl strip charged with 150V DC bias with respect to the static
receiver electrode. The polyvinyl strip is glued around the central
post of the TO, except for the section near the brass disc which is
suspended inside the sample space, in parallel to the static electrode,
at a distance of several hundred microns. To make the device more
sensitive, the suspended section of the movable electrode is not under
any tension. Motion of the charged movable electrode will induce a
current in the receiver electrode which would be detected by a current
preamplifier. During the experiment, the gap between the two electrodes
is also filled with solid He, so that the effective mechanical
compliance of the microphone is similar to that of solid He. Our
working model is illustrated schematically in Fig.
\ref{fig:working-model}. We assume that at some distance $D$ from the
wall of the annular channel there is a grain boundary separating the
solid moving with the wall and the decoupled solid. Both electrodes of
the microphone are embedded inside the solid moving with the wall.
Lateral motion of the decoupled solid along this grain boundary will
generate a time dependent force between the two surfaces resulting from
the periodicity of the lattice potential. The strain generated by this
force can be detected by the microphone. The extent of the lateral
relative motion at the interface is controlled by the oscillation
amplitude of the TO. This amplitude is small, typically a few lattice
constants. On that scale, there is a reasonable chance that the
interface will contain no defects. In this case, this motion will
induce a periodic vibrating strain at the microphone with a frequency
$f=V_{\mathrm{rel}}/d$, where $V_{\mathrm{rel}}$ is the relative
lateral speed between the wall and the decoupled solid and $d$ is the
spacing between nearest rows of atoms of the moving solid. The time
dependent signal of the microphone can be Fourier transformed to
determine this frequency, and thus yield information about the
interatomic distance and the flow speed.	

The expected strain amplitude near the movable electrode should be much
less than a lattice constant of solid He. Modelling the microphone as a
capacitor, its measured noise floor is low enough to detect vibrations
down to the 10$^{-11}$\,m range (a few \% of a lattice constant). The
bandwidth is 250\,kHz, limited by the current amplifier. Such bandwidth
is required since for a typical relative speed of several $\mu$m/sec,
the expected frequency $f=V_{\mathrm{rel}}/d$  is tens of kHz, the
exact value depending on the interatomic distance $d$.

The principle of the method proposed here is similar to Atomic Force
Microscopy (AFM), which unfortunately cannot be applied to solid He,
since the He-He atomic bond is so weak that the interaction with any
AFM tip would destroy the surface of the He crystal. In our case, the
movable electrode, effectively the ``AFM tip'', is coated by solid He,
as are all the internal surfaces of the sample cell. Thus, its
interaction with crystal grains moving at the interface (Fig.
\ref{fig:working-model}) should be that of solid He with solid He,
which at small stress as we use is non-destructive. One additional
inconvenience is that unlike with an AFM, where one scans the sample
until a clean surface is found, we have to grow a new crystal to try
again. Growth and characterization of a polycrystal typically takes
more than a week.

A priori, there is a significant probability that the experiment would
yield a null result even in the presence of flow of the solid relative
to the wall. Referring again to Figure \ref{fig:working-model}, if the
distance $D$ between the microphone and the spontaneously formed grain
boundary is large, the vibrations at the microphone may be too weak to
detect. Second, the lateral size of our microphone is about a mm, while
the typical size of the crystal grains making up the polycrystal is a
fraction of a mm \cite{a018}. Vibrations originating from different grains are not coherent,
and so the total signal of the microphone would average to zero. Based on the analogy with an AFM, it is
likely that a non-zero signal would be detected only
if some small part of the microphone, smaller than a single grain (0.1\,mm-0.3\,mm), is close enough to
the grain boundary.  Our
samples consist of similar, but not identical single crystals. Our hope
was that some of these crystals will satisfy the conditions listed above.

\begin{figure} % Figure 5 response-spectra
  \includegraphics[width=1\textwidth]{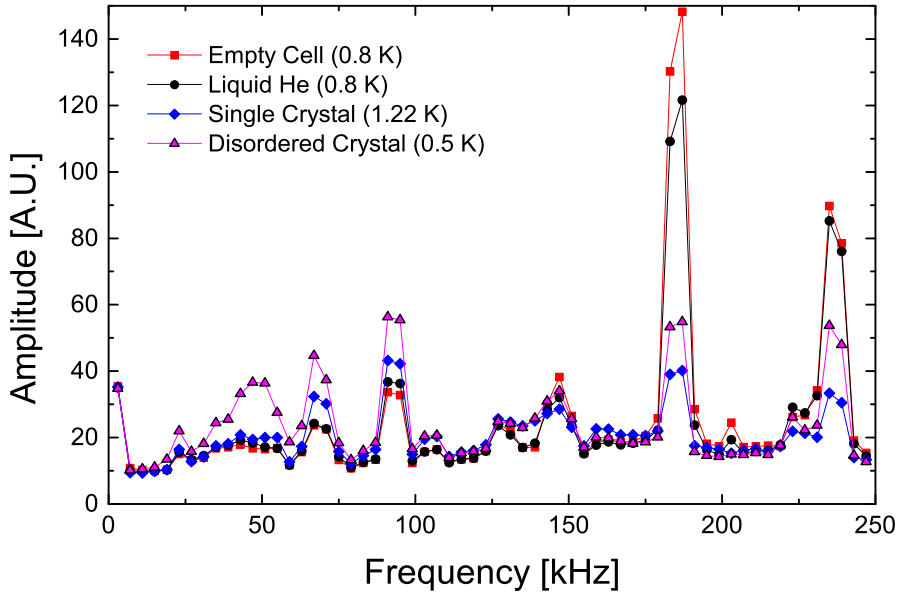}

	\caption{
		Typical response spectra of the microphone signal under different conditions. The black arrow points out a set of data taken with a disordered crystal in the cell, at a wall speed of $5.6\,\mu$m/sec.
	}
	\label{fig:response-spectra}
\end{figure}

Detecting the vibrations is further complicated by the fact that they
originate at the atomic scale and so are not phase correlated with the
oscillatory motion of the TO. Therefore, simply averaging the signal of
the microphone over many cycles of the TO sums to zero. Instead, we
averaged the absolute value of the Fourier transform of the microphone
signal. This procedure makes the data insensitive to the phase of the
signal. However, one problem which arises in this case is that the
random noise does not cancel. Therefore, the power spectra contain
contributions of both signal and noise. In order to extract meaningful
results, one first needs to understand the response and noise
characteristics of the microphone. We studied the background noise by
analyzing spectra obtained with the cell either empty or filled with
liquid. A typical set of spectra acquired under different conditions is
shown in Figure \ref{fig:response-spectra}. We first consider the
spectrum of an empty cell (red symbols). One can see that the noise
spectrum is basically flat with several resonance-like features. We
believe that these are acoustic ``plate resonances'' of the polyvinyl
electrode, which like everything else freezes and becomes rigid at low
temperatures. These resonances are excited by the ambient broadband
electrical field noise which is always present. Since the movable
electrode is biased at 150\,V, the electrical charge on it would react to
electric fields. Filling the cell with liquid He (green symbols)
reduces the amplitude of these resonances, but does not affect their
frequency. To see if these resonances couple to sound waves in He, we
changed the pressure of the liquid between SVP and 25 bars. The speed
of sound in liquid He changes from 237\,m/s at SVP to 365\,m/s at 25\,bar,
but there was no change in the frequencies of the resonances. The only
effect seen in Fig. \ref{fig:response-spectra} is that the amplitude of
the resonances is reduced. Hence, the decrease of the amplitude of the
resonances in the presence of liquid is due to mass loading of the
movable electrode, and there is no coupling of the microphone to sound
waves in He.

Filling the cell with a single crystal (blue symbols) does not produce
any additional features in the spectrum in Fig.
\ref{fig:response-spectra}. However, it reduces the amplitude of the
resonances sharply. This is to be expected since single crystals are
rigid and immobilize the electrode more effectively than the liquid.
Finally, these plate resonances showed no dependence on temperature or on the speed of
the TO. The only effect the TO has on the spectra is one point near
zero frequency in Fig. \ref{fig:response-spectra}, which results from
feedthrough at the drive frequency (360\,Hz), and is the same for all
data sets.

Having understood the background, we next look for a signal arising
from a moving disordered crystal. Such a spectrum is shown in Figure
\ref{fig:response-spectra} by violet triangles. We first notice that the
intensity of the plate resonances of the electrode is somewhat higher
with a disordered crystal than with a single crystal. This is
reasonable, since the disordered crystal is not as rigid as a single
crystal and so is less effective in immobilizing the electrode. More
importantly, with a disordered crystal we see an additional feature
which was not there in the single crystal or empty cell spectra, namely
a broad peak around 50\,kHz. When referred to the input, the magnitude of
this peak represents motion of the electrode in the 10$^{-11}$\,m range,
which is the correct magnitude for the signal we expected. The
frequency at which the peak appears is also in the region we expected.
The spectra acquired with disordered crystals were the only ones which
depended on the speed of the TO. To extract useful information, we
divided each power spectrum obtained with a disordered crystal by that
of its parent single crystal which is static. This procedure allowed us
to remove most of the frequency dependence of the background by using
the single crystal spectrum as an effective gain function. The features
which emerge under this procedure are the excess vibrations due to the
motion of the polycrystal relative to a static single crystal.

\section{Results and Discussion}
\label{results}

In total, we grew and measured 15 hcp oriented polycrystals. Most of
the parent single crystals were grown on the melting curve between 1.2\,K
and 1.4\,K. Upon disordering, all the polycrystals showed a similar
decoupling mass fraction. Excess vibrations of the kind shown in Fig.
\ref{fig:response-spectra} as violet triangles were detected in 3
polycrystals. This indicates that although all the polycrystals showed
the same mass decoupling so that all of them move relatively to the
wall, the stringent conditions for detection of the motion by the
microphone are satisfied only by some of them. In addition, such
vibrations were seen only at T=0.5\,K, our lowest temperature, where the
dissipation of the TO was close to that of an empty cell.

\begin{figure} % Figure 6 - single
  \includegraphics[width=1\textwidth]{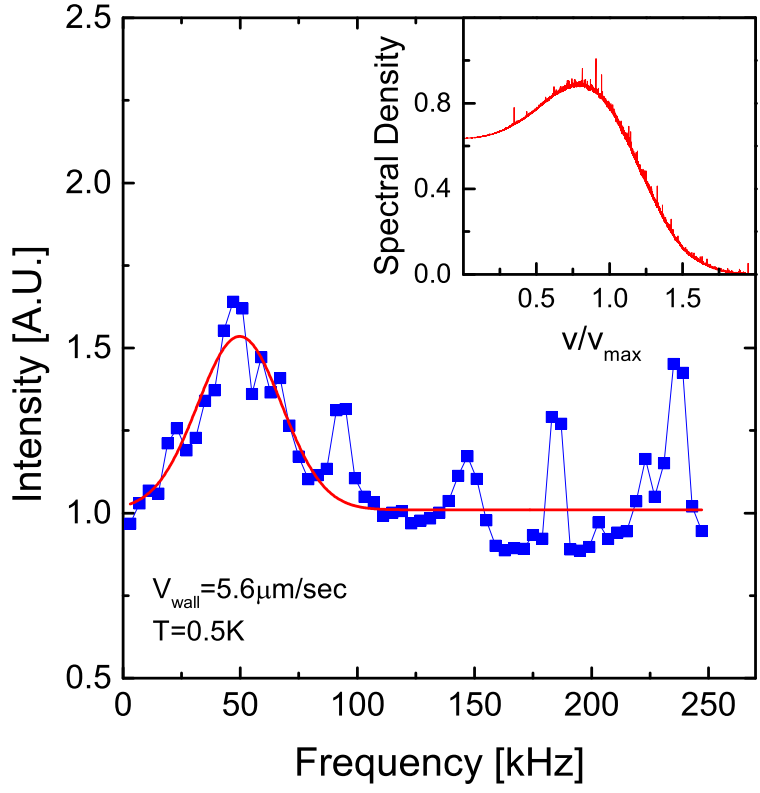}

	\caption{
		Typical power spectrum of vibrations at 0.5\,K. The data are of the disordered crystal shown in Fig. \ref{fig:response-spectra}. The features around 80,150,180 and 230\,kHz are part of the background. Solid line is a fit. The inset shows a simulated spectrum in the presence of noise of the TO.
	}
	\label{fig:single}
\end{figure}

The data shown in Fig. \ref{fig:single} represent a typical spectrum of
excess vibrations of a disordered crystal relative to its parent single
crystal. The main feature is a broad peak around 50\,kHz. To understand
the origin of this peak, we consider the case where the moving surface is
free of defects and well defined vibrations exist. During each cycle of the TO,
the speed of the solid relative to the wall changes between zero and
some maximum value $V_{\mathrm{max}}$ as
$V(t)=V_{\mathrm{max}}\cos\left(\omega_{TO}t\right)$. The instantaneous
frequency of vibrations $f$ would therefore change between zero and
$f=V_\mathrm{max}/d$. The intensity of vibrations at a given frequency
$f$ in the spectrum depends on the amount of time the TO spends at a
particular speed $V$ during each cycle. We therefore calculated the
probability of finding the TO moving at a certain velocity at some
random time. This probability is proportional to the time $dt$ spent at
each velocity between $V$ and $V+dV$. It can be calculated from the
derivative of the inverse time dependence of the speed,
$t(V)\propto\arccos(V/V_{\mathrm{max}})$. We find that the resulting
probability density function $p$ depends on speed as
$dt/dV\propto p(V)=2/(\pi\sqrt{1-(V/V_{\mathrm{max}})^2})$. Consequently, the power %Added \dt/dV\propto (EL)
spectrum of the vibrations shows a divergence at a frequency $f=V_{\max} /d$. 		%changed "peak" to "divergence" (EL)
In the actual data we expect that finite frequency resolution and noise
will wash out the singularity, but simulations done in the presence of
noise indicate that a peak still exists. The result is shown as the
inset in Figure \ref{fig:single}. We expected the peak frequency $f$ to
depend on friction. For a given $V_{\mathrm{wall}}$, the highest value
of $f$ will be measured if the friction is zero. In this case the
decoupled solid does not move at all, namely $V_{\max} =
V_\mathrm{wall}$, and so $f=V_{\mathrm{\mathrm{wall}}} /d$. In the
limit of high friction, the wall and the solid move together and $f=0$.
With finite friction, for example of the stick-slip type, there may be
no well defined frequency associated with the relative motion.

Observing a well defined peak of excess vibrations at some frequency
confirms that (a) the solid inside the cell is indeed moving relative to the
wall. (b)  the surface of the moving grains is free of defects on the
scale of the motion.

\begin{figure} % Figure 7 - famous-all
  \includegraphics[width=1\textwidth]{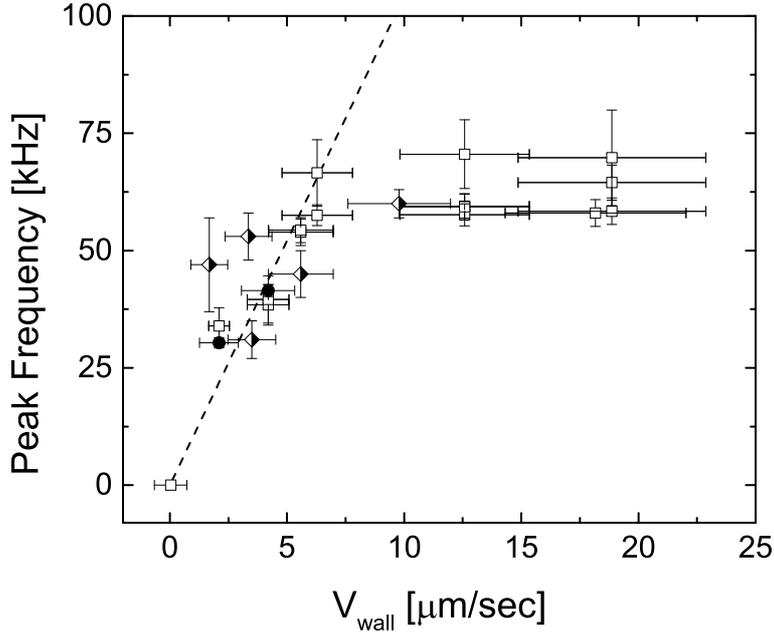}

	\caption{
		Peak frequencies of vibrations at 0.5\,K vs. wall speed. Various symbols denote data obtained from 3 different polycrystals. Dashed line is a guide to the eye. The error bars combine the width of each vibration peak (as in Fig. \ref{fig:single}) and the uncertainty in determining the absolute value of the speed of the TO.
	}
	\label{fig:famous-all}
\end{figure}

\begin{figure} % Figure 8 - famous-best
  \includegraphics[width=1\textwidth]{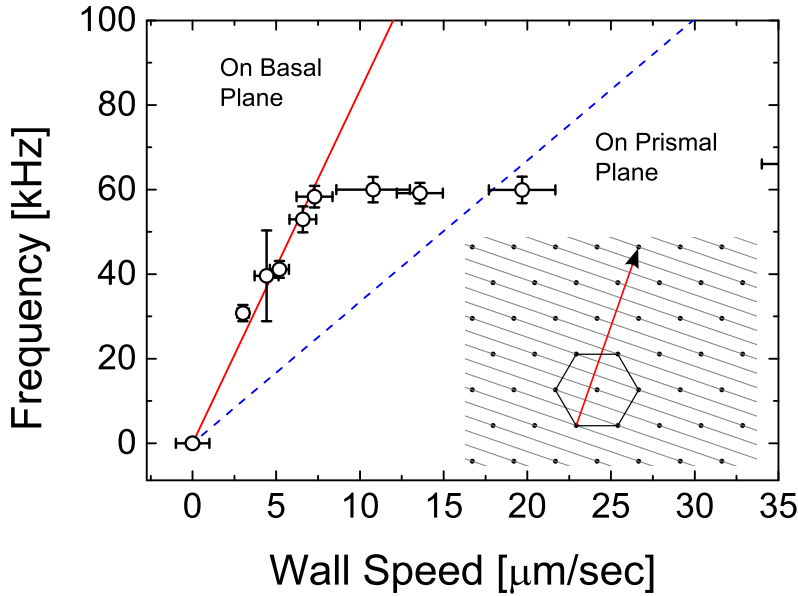}

	\caption{
		Averaged peak frequencies at 0.5\,K plotted vs. wall speed. The lines show the relation expected in the case of zero friction. Solid red line: the surface of the moving solid is the (0001) (basal) plane, in the direction shown in the inset. For comparison, we show the relation expected for motion in the [0001] direction in the $(010\bar{1})$ (c-a) plane as the dashed line. These two planes have the highest atomic density.
	}	%Changed (010-1) to (010\bar{1})
	\label{fig:famous-best}
\end{figure}

In Fig. \ref{fig:famous-all} we show all the peak frequencies taken
from data sets such as shown in Fig. \ref{fig:single}, obtained from 3
polycrystals. The data are plotted vs. wall speed. At small speeds, the
peak frequencies increase with $V_{\mathrm{wall}}$. At higher speeds,
the peak frequencies level off. In Fig. \ref{fig:famous-best} we show
the same data after averaging all the points taken with the same wall
speed.  Evidently, at small speeds the peak frequency increases
linearly with $V_{\mathrm{wall}}$. The lines in the figure show the
$f$-$V_{\mathrm{wall}}$  relation for the case where the friction is
zero ($f=V_{\mathrm{wall}} /d$). Taking the friction to be zero is
consistent with the overall zero dissipation which we measured at 0.5\,K.
The two lines in Fig. \ref{fig:famous-best} differ by the value of $d$,
the distance between nearest rows of atoms. The value of $d$ which fits
the data is for the surface of the moving solid being (0001), the basal
plane of the hcp structure, with the in-plane direction of motion shown
in the inset. This result suggests that under the shear stress applied
by the TO, the crystalline grains near the wall reorient themselves
with their (0001) planes parallel to the wall, so their $c$ axis is
horizontal. This re-orientation takes place during the relaxation
process shown in Fig. \ref{fig:break}.
%In previous experiments\cite{a020} we found that typical dimensions of
%a grain is $\\sim$0.1 mm. With the diameter of the annular sample
%space of 14 mm, there should be several hundred such grains around the
%circumference of the channel, with misorientation angle of less than
%1$^{0}$ on average.
Finally, we remark that (0001) is the only natural slip
plane \cite{a008,a010} for hcp $^4$He.

It is interesting that we were able to observe well defined vibrations
only at 0.5\,K, where the overall dissipation inside the solid tends to
zero, but not at higher temperatures. One would think that with small
enough friction it would still be possible to observe vibrations,
perhaps at a lower frequency or weaker in intensity, but this was not
the case. It may be that at temperatures above 0.5\,K, where we detect a
finite dissipation, the relative motion at the interface is not regular
enough to produce vibration we can detect. This could happen for example if the friction at temperatures above 0.5\,K is of the
slip-stick type.		%removed a redundant space (EL)

Referring again to Fig. \ref{fig:famous-best}, for $V_{\mathrm{wall}}
\le 7\,\mu$m/sec, the peak frequency saturates at $\sim$60\,kHz. This
levelling off suggests that the conditions of zero friction which
account for the linear part no longer apply. However, well defined
vibrations are still observed. Simulations done with two masses coupled
by a friction force proportional to velocity indicate that the
frequency of vibrations, proportional to
$(V_{\mathrm{wall}}-V_{\mathrm{solid}})$, should gradually decrease to %minus sign did not appear in compilation. Fixed (EL)
zero as $V_{\mathrm{wall}}$ increases instead of staying constant. We
also considered an alternate possibility, that the frequency in Fig.
\ref{fig:famous-best} saturates due to presence of defects at the
interface which limit the amplitude of frictionless motion to some
finite value. We assume that at higher amplitudes, the friction is
finite.  Calculations done with this scenario show that the peak frequency 	%I don't think the word simulation applies here,
continues to change with wall speed, and does not saturate. Hence,
those simple scenarios cannot explain the data.

One possibility of understanding the presence of a low friction state follows from the results of
Path Integral Monte Carlo (PIMC) simulations of grain boundaries in hcp $^4$He by Pollet, et al.\cite{a020}.  Pollet, et al. showed that in equilibrium, the interfaces in most grain boundaries are fluid.  The initial configuration of their simulation\cite{a020} is that of two solid grains of different orientations in contact along a GB. Once the simulation began, 2-3 atomic layers at the interface spontaneously became fluid. The reason for this is thermodynamic- the presence of fluid lowers the interfacial energy. In addition, the orientation of the GB spontaneously changed from the initial one. These authors were able to conclude that a fluid layer appears on most GB's, but not on all of them. No less important, they found that the fluid layer is not an usual bulk liquid - it is metastable, in the sense that it appears only at a GB. These PIMC results were largely confirmed experimentally in inelastic neutron scattering on solid He contained inside a porous material\cite{a021}. This experiment\cite{a021} detected superfluid-like excitations inside the solid. In a porous material with a large surface area, the contribution of the He near surfaces to the scattering intensity is substantial, and so the experiment was able to detect them. Microscopic theory calculations presented in the same paper show that the presence of a 1-2 atomic layers of superfluid at the interfaces would explain these results.
In the present context, a fluid layer which spontaneously appears at the internal interfaces inside the TO would act as a natural lubricant which would allow the grains to ''slide'' against each other. Its spontaneous appearance at grain boundaries might explain why ''mass decoupling'' is observed as soon as the single crystal is disordered (Figs. \ref{fig:break} and \ref{fig:qfactor})and grain boundaries are created. Since the PIMC simulations show that the fluid is present at most GB's, then the GB's inside our TO would not have to be identically oriented around the circumference of the annular channel. Finally, these simulations\cite{a020} show that the grain boundaries become superfluid around 0.5K. This may be related to the observation that the measured internal dissipation of our TO goes to zero at 0.5K.

In conclusion, we developed a new technique to characterize the
interface between He crystallites as they move past each other. We
confirmed that this relative motion indeed exists and the surface of
the moving crystallites is the basal plane of the hcp structure. These
findings are essential for future studies of friction. Our results are
compatible with the notion that at small enough speeds, the motion of
grains with the (0001) surface is frictionless at T=0.5\,K.

%as required. Don't forget to give each section
%and subsection a unique label (see Sect.~\ref{sec:1}).
%\paragraph{Paragraph headings} Use paragraph headings as needed.
%\begin{equation}
%\end{equation}

%\begin{figure*}
%  \includegraphics[width=0.75\textwidth]{example.eps}
%\end{figure*}
%

\begin{acknowledgements}
We acknowledge useful discussions with A. Auerbach and B. Svistunov.  We thank A. Danzig, S. Hoida and L. Yumin for their assistance. This work was supported by the Israel Science Foundation and by the Technion Fund for Research.
\end{acknowledgements}

\subsection{Appendix - Finite Element Analysis}

Maris and Balibar\cite{a016} were the first to point out that the resonance frequency of the TO can be affected by the changes of the shear modulus of the solid He. Specifically, they addressed the low temperature anomalies reported by Kim and Chan\cite{a003} around 0.2K. The shear modulus of the solid changes significantly around this temperature\cite{a007}. Over the temperature range of our experiment (0.5K-2K), the shear modulus changes by about 30\%, as reported by Paalanen, et al.\cite{a011}. The shear modulus increases gradually towards low temperatures. To see how this affects our experiment, we performed a Finite Element Analysis of our TO. We simulated two different situations. First, we checked the decrease of the resonant period resulting from the increase of the shear modulus as the temperature is reduced from 2K to 0.5K.  The simulated decrease in this case is about 7\,nsec, while our observed decrease of the period (due to what we call ''mass decoupling'')is typically between 0.6\,$\mu$sec and 1\,$\mu$sec, a hundred times larger. In addition, the elastic properties of the solid change smoothly as the temperature decreases from 2K to 0.5K, while in our experiment the majority of the period shift occurs over a 10\,mK interval near the temperature at which the crystal was grown. We conclude that our observations cannot be explained by the temperature dependence of the shear modulus of solid He.

\begin{figure} % Figure 9 FEA
	\begin{center}
		\includegraphics[width=1\textwidth]{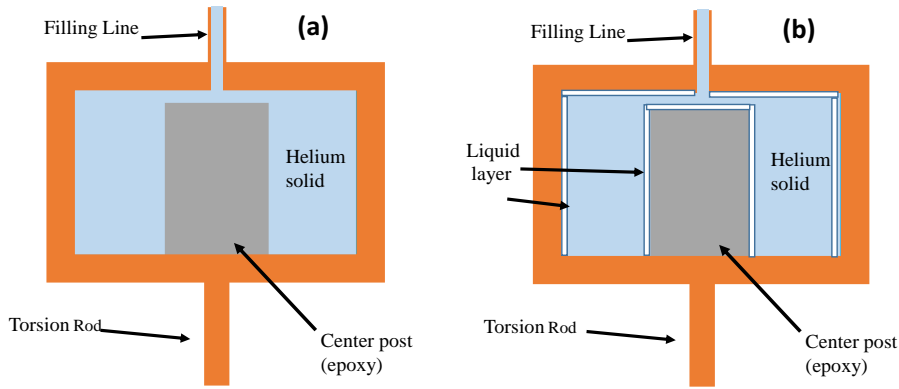}
	\end{center}
	\caption{
 Schematic illustration of a typical simulated configuration of having both liquid and solid inside the cell. Panel (a) shows a schematic cross section of our TO filled with solid He. Panel (b) show a situation where the solid He is attached to the wall of the cell only at the bottom, while a thin layer of liquid separates the solid from all the other walls.}
	\label{fig:FEA}
\end{figure}

Another scenario which we checked in conjunction with the decrease of the resonant period which we see (Figs. \ref{fig:break} and \ref{fig:qfactor}) is the following:  Suppose that at the end of growth, a small amount of liquid remains trapped in the cell so that the solid is not fully coupled with the walls of the TO. Once the cell is cooled, this liquid freezes and the solid He becomes coupled to the walls. This coupling might perhaps make the TO ''stiffer'', increase the effective torsion constant and decrease the period. To check this scenario, we extended the finite element analysis to simulate the TO in several configurations where some fluid is present in the cell in different locations. To illustrate what we did, we show one such configuration in the figure \ref{fig:FEA}
In this particular configuration the coupling of the solid He to the walls is minimal, and so the presence of the liquid had the largest effect. The simulated period of the TO was compared with that of a cell filled with solid. Since the amount of solid He (and moment of inertia) of the cell containing fluid is different than that of a cell filled with solid, for each configuration we carried out a series of FEA simulations with different amounts of liquid and extrapolated to zero liquid content. The extrapolated value was used in the comparison.

For the configuration shown in figure \ref{fig:FEA}, in the limit of zero liquid content, the simulated period shift is (+)71 nsec. This should be compared to the observed shift of (-)0.6\,$\mu$sec to (-)1\,$\mu$sec in different crystals and TO’s. The difference in sign results from the fact that in the accelerated frame of the TO, a fictitious D’Alambert force acts on the solid He during the oscillation. Due to its small resistance to shear, in the presence of liquid the solid He is free to deform slightly under this force and so effectively it appears to oscillate in antiphase to the motion of the TO. This type of response increases the effective torsion constant and so the effective period is smaller. When the cell is filled with solid, this internal oscillation does not take place, the additional contribution of the solid He to the effective torsion constant is now smaller  and so the period increases. This gives the (+) sign.
In all cases, we found that the simulated period shift was of the opposite sign and at least an order of magnitude smaller than what we observed in the experiment. We conclude that the presence of liquid cannot explain our observations.

\bibliographystyle{spphys}        % APS-like style for physics
%\bibliography{frictionless_jltp_revised}  % name your BibTeX data base

\begin{thebibliography}{}
\bibitem{a001}
O. Pelleg, M. Shay, S.G. Lipson, E. Polturak, J. Bossy, J.C.  Marmeggi, K. Horibe, E.   Farhi, and A. Stunault, Phys. Rev. {\bf{B73}},
,024301 (2006).
\bibitem{a002}
C.A. Burns, N. Mulders, L. Lurio, M.H.W. Chan, A. Said, C.N. Kodituwakku, and P.M. Platzman, Phys. Rev. {\bf{B78}}, 224305 (2008).
\bibitem{a003}
E. Kim and M.H.W. Chan, Nature {\bf{427}}, 225 (2004).
\bibitem{a004}
D. Y. Kim and M. H. W. Chan, Phys. Rev. Lett. {\bf{109}}, 155301 (2012).
\bibitem{a005}
Xiao Mi, Anna Eyal, A. V. Talanov, and J. D. Reppy, arXiv:1407.1515
\bibitem{a006}
G. Nichols, M. Poole, J. Nyeki, J. Saunders, and B. Cowan, arXiv:1311.3110
\bibitem{a007}
J. Day, and J. Beamish, Nature {\bf{450}}, 853 (2007).
\bibitem{a008}
A. Haziot, A. D. Fefferman, F. Souris, J. R. Beamish, H. J. Maris, and S. Balibar, Phys. Rev. B{\bf{88}}, 014106 (2013)
\bibitem{a009}
X. Rojas, A. Haziot, V. Bapst, H. J.  Maris, and S. Balibar, Phys Rev. Lett. {\bf{105}},
145302-5 (2010).
\bibitem{a010}
A. Haziot, X. Rojas, A. D. Fefferman, J. R. Beamish, and S. Balibar, Phys. Rev.
 Lett. {\bf{110}}, 035301 (2013)
\bibitem{a011}
M. A. Paalanen, D. J. Bishop, and H. W. Dail,  Phys. Rev. Lett. {\bf{46}}, 664 (1981).
\bibitem{a012}
F. Tsuruoka and Y. Hiki, Phys. Rev. B{\bf{20}}, 2702, (1979).
\bibitem{a013}
J. B. Pendry, New J. Phys. {\bf{12}}, 03302814, (2010).
\bibitem{a014}
V.L. Popov, Phys. Rev. Lett. {\bf{83}}, 1632, (1999).
\bibitem{a015}
A. Eyal, E. Livne, and E. Polturak, to be published.
\bibitem{a016}
H. Maris and S. Balibar, Jour. Low Temp. Phys. {\bf{162}}, 12 (2011).
\bibitem{a017}
A. Eyal, O. Pelleg, L. Embon, and E. Polturak, Phys. Rev. Lett. {\bf{105}}, 025301 (2010).
\bibitem{a018}
A. Eyal, and E. Polturak, Jour. Low Temp. Phys. {\bf{163}}, 262 (2011).
\bibitem{a019}
A. Eyal and E. Polturak, J. Low Temp. Phys. {\bf{168}}, 117 (2012).
\bibitem{a020}
L. Pollet, M. Boninsegni, A. Kuklov, N. Prokof’ev, B. Svistunov, and M. Troyer,
Phys. Rev. Lett. {\bf{98}}, 135301 (2007).
\bibitem{a021}
H. Lauter, V. Apaja, I. Kalinin, E. Kats, M. Koza, E. Krotscheck, V. V. Lauter, A. V.
Puchkov, Phys. Rev. Lett. {\bf{107}}, 265301 (2011).
% etc
\end{thebibliography}


\end{document}